\documentclass[twocolumn]{aastex631}

\begin{document}

\title{Peculiar Velocity Reconstruction From Simulations and Observations Using Deep Learning Algorithms}

\correspondingauthor{Yuyu Wang}
\email{yuyuwang@sjtu.edu.cn}

\author[0000-0002-0245-8547]{Yuyu Wang}
\affiliation{Department of Astronomy, School of Physics and Astronomy, and Shanghai Key Laboratory for Particle Physics and Cosmology, Shanghai Jiao Tong University, Shanghai, 200240, China.}

\author[0000-0003-3997-4606]{Xiaohu Yang}
\affiliation{Tsung-Dao Lee Institute, and Key Laboratory for Particle Physics, Astrophysics and Cosmology,  Ministry of Education, Shanghai Jiao Tong University, Shanghai 201210, China}
\affiliation{Department of Astronomy, School of Physics and Astronomy, and Shanghai Key Laboratory for Particle Physics and Cosmology, Shanghai Jiao Tong University, Shanghai, 200240, China.}

\begin{abstract}
In this paper, we introduce a Unet model of deep learning algorithms for reconstructions of the 3D peculiar velocity field, which simplifies the reconstruction process with enhanced precision. We test the adaptability of the Unet model with simulation data under more realistic conditions, including the redshift space distortion (RSD) effect and halo mass threshold. Our results show that the Unet model outperforms the analytical method that runs under ideal conditions, with a 16\% improvement in precision, 13\% in residuals, 18\% in correlation coefficient and 27\% in average coherence. The deep learning algorithm exhibits exceptional capacities to capture velocity features in non-linear regions and substantially improve reconstruction precision in boundary regions. We then apply the Unet model trained under SDSS observational conditions to the SDSS DR7 data for observational 3D peculiar velocity reconstructions.
\end{abstract}


\section{Introduction} \label{sec:intro}

The galaxy distribution and peculiar velocity are two powerful engines in large-scale structure studies. While the galaxy distribution has been explored and implemented in many studies \cite[e.g.][]{KerSzaSza2000, VarBauHam2013, WanBruDol2013, CacVosMor2013, ManSloBal2013, KeiKurLin2019, XuLiZha2023, LiuZhaLi2023}, which is benefited from large redshift surveys \cite[e.g.][]{SDSS2000, JonReaSau2009, HucMacMas2012, FlaBeb2014}, the study of peculiar velocity is still limited by its survey size due to difficulties in distance measurements. Unlike the galaxy distribution, which is a biased tracer of mass distribution, peculiar velocity is an unbiased tracer and can provide an independent perspective to boost cosmological studies \cite[e.g.][]{WatFel2015a, HelNusFei2017, WanRooFel2018, dupCouKub2019, WanPeeFel2021, TurBlaRug2021, WatAllBra2023, BlaTur2023, BouColSai2023, TurBlaRug2023, LaiHowDav2023}.

Under current observation conditions, peculiar velocity surveys are limited to a very tiny redshift region due to the fatal uncertainty in peculiar velocity estimation from distance modulus at high redshifts. The CosmicFlow-4 \cite[CF4,][]{TulKouCou2023} dataset, which includes 55,877 galaxies up to $z\sim 0.1$, is currently the largest peculiar velocity dataset that compiles distance and combines velocities from many surveys such as the 6-degree Field Galaxy Survey \cite[6dFGSv,][]{SprMagCol2014}. On the other hand, redshift surveys such as the Dark Energy Spectroscopic Instrument \cite[DESI,][]{FlaBeb2014} Survey have been targeting tens of millions of objects with redshift reaches 2.1. The fast development of redshift surveys brings opportunities to peculiar velocity studies. Instead of measuring the peculiar velocity of each galaxy by its distance modulus, one can reconstruct velocity fields from over-density fields calculated by galaxy redshift surveys, which could effectively expand the velocity studies to larger redshift regions.

Reconstructing an unbiased tracer (velocity field) from a biased tracer (galaxy density field) can be very complicated. First, the redshift space distortion (RSD) is inevitable in calculating the over-density field from the redshift distribution of galaxies. The RSD effect in the velocity reconstruction can be improved by increasing iteration times or enlarging smooth width, though it complicates the calculation. Second, the over-density field calculated from galaxy distribution is biased since the galaxy is a biased tracer of mass distribution. It can be improved by introducing halo mass bias and density reconstruction into the velocity reconstruction process. Many approaches have been studied to improve the velocity reconstruction result and process \cite[e.g.][]{CroSilZuk2010, CouHofTul2012, WanMoYan2012, Lavaux2016, KesNus2017, SorHofGot2017, YuZhu2019, ZhuWhiFer2020, BorLavHud2022, ValLibHof2023, BayModFer2023, TurBla2023, HofValLib2023}. However, the analytical velocity reconstruction process is still complicated and computationally expensive. The advancement of machine learning algorithms has inspired a new way to achieve velocity reconstruction, which is faster, more accurate, and more efficient.

Machine learning algorithms are designed to perform complex analyses in a data-driven manner without explicit programming of the physical phenomena. One subset of these algorithms is the deep learning algorithm, which are extremely flexible statistical models that can automatically extract relevant features with a large number of trainable parameters optimized from abundant amounts of data. In recent years, the utilization of machine learning algorithms in astrophysics has rapidly increased, including cosmological value predictions \cite[e.g.][]{WanRamSal2021, ChuMah2023}, galaxy morphological classification \cite[e.g.][]{DomMarBer2022, LiXuLi2023}, and more \cite[see][for a recent overview]{Baron2019}. Many studies have employed deep learning algorithm in velocity reconstruction \cite[e.g.][]{WuZhaPan2021, QinParHon2023, WuXiaXia2023, GanLilNus2023}. Diverging from theory-driven algorithms, deep learning operates as a data-driven approach, with its effectiveness directly tied to the quality and quantity of the training dataset and its performance evaluated by the employed loss function. Complex issues encountered in theory-driven algorithms, such as noise processing and data preprocessing, can be simplified within deep learning algorithms by integrating them into the training data without additional processing. The loss function quantifies the difference between predictions and expectations (training outputs). For instance, in regression problems, the widely adopted mean square error loss function measures the squared difference between predictions and expectations. Ideally, a well-designed deep learning model yields more precise predictions when armed with an appropriate loss function and a large training dataset enriched with comprehensive and diverse information, as discussed in \citep{GanLilNus2023}.

In this paper, we present a deep neural network model that reconstructs the 3-dimensional peculiar velocity field from the galaxy redshift distribution, and  utilize the model to analyze both simulation and observation data. The model is based on a modified U-net architecture, which is a fully convolutional neural network originally designed for biomedical image segmentation by the Computer Science Department of the University of Freiburg \citep{RonFisBro2015}. In section~\ref{sec:method}, we detail the analytical method and deep learning algorithms of the velocity reconstruction. In section~\ref{sec:simul}, we introduce the simulation data used for model training and validation. In section~\ref{sec:training}, we elucidate the training strategy employed for deep learning model. In section~\ref{sec:results}, we show the performance of the deep learning model and compare it with ideal analytical results. In section~\ref{sec:observ}, we apply the model to observation data. Finally, in section~\ref{sec:conclusion}, we conclude this paper.

\section{Methods} \label{sec:method}

\subsection{Analytical Method} \label{sec:FFT}

In linear perturbation theory, the equations of motion (including Continuity, Euler, and Poisson equation) that describe the relation between density and velocity are derived by introducing small perturbations to the equations of an ideal fluid. By applying Fourier transformation to these equations, we obtain:
\begin{equation} \label{eq:FFT}
v(\textbf{k}^\alpha)=iHaf\frac{\textbf{k}^\alpha}{\textbf{k}^2}\delta(\textbf{k}) ,
\end{equation}
where $v(\textbf{k})$ represents the velocity field in Fourier space, $\delta(\textbf{k})$ indicates the over-density filed in Frourier space, $H$ is the Hubble constant, $a$ expresses the scale factor, $f$ describes the growth rate, and $\textbf{k}^\alpha$ indicates the component along $\alpha$th (x, y and z) direction.

Due to observational conditions and the existence of dark matter, the observable mass is not an unbiased cosmological tracer. As a result, the velocity field reconstructed from halos (galaxies or groups) exhibits bias, which can be improved by incorporating the average bias of halos, denoted as $b_{hm}$:
\begin{equation} \label{eq:halo_bias}
b_{hm}=\frac{\int_{M_{th}}^\infty Mb_h(M)n(M)dM}{\int_{M_{th}}^\infty Mn(M)dM} ,
\end{equation}
where $M_{th}$ represents the minimum mass threshold of halos, $b_h(M)$ indicates the halo bias function, and $n(M)$ describes the halo mass function. Under these circumstances, equation~\ref{eq:FFT} can be written as:
\begin{equation} \label{eq:FFT_new}
v(\textbf{k}^\alpha)=i\frac{Haf}{b_{hm}}\frac{\textbf{k}^\alpha}{\textbf{k}^2}\delta_h(\textbf{k}) ,
\end{equation}
where $\delta_h$ indicates the over-density field calculated from halos.

Ideally, velocity reconstruction requires over-density in real space to avoid the RSD effect. However, the measurement uncertainty and sample size of distance surveys are inferior to those of redshift surveys. Consequently, practical velocity reconstruction employs the over-density field in redshift space, even though it introduces RSD, which affects the reconstruction results. However, the RSD effect can be mitigated by increasing the field gridding size (or Gaussian smoothing width) or by correcting the RSD effect through iterations \cite[e.g.][]{WanMoYan2012, HolHud2021}.

In this paper, we apply the analytical method under ideal conditions, which is discussed in detail in section~\ref{sec:simul}, to compare its performance with that of the deep learning algorithm.

\subsection{Deep Learning Algorithm} \label{sec:DL}

\begin{figure*}
\centering
\includegraphics[width=16cm]{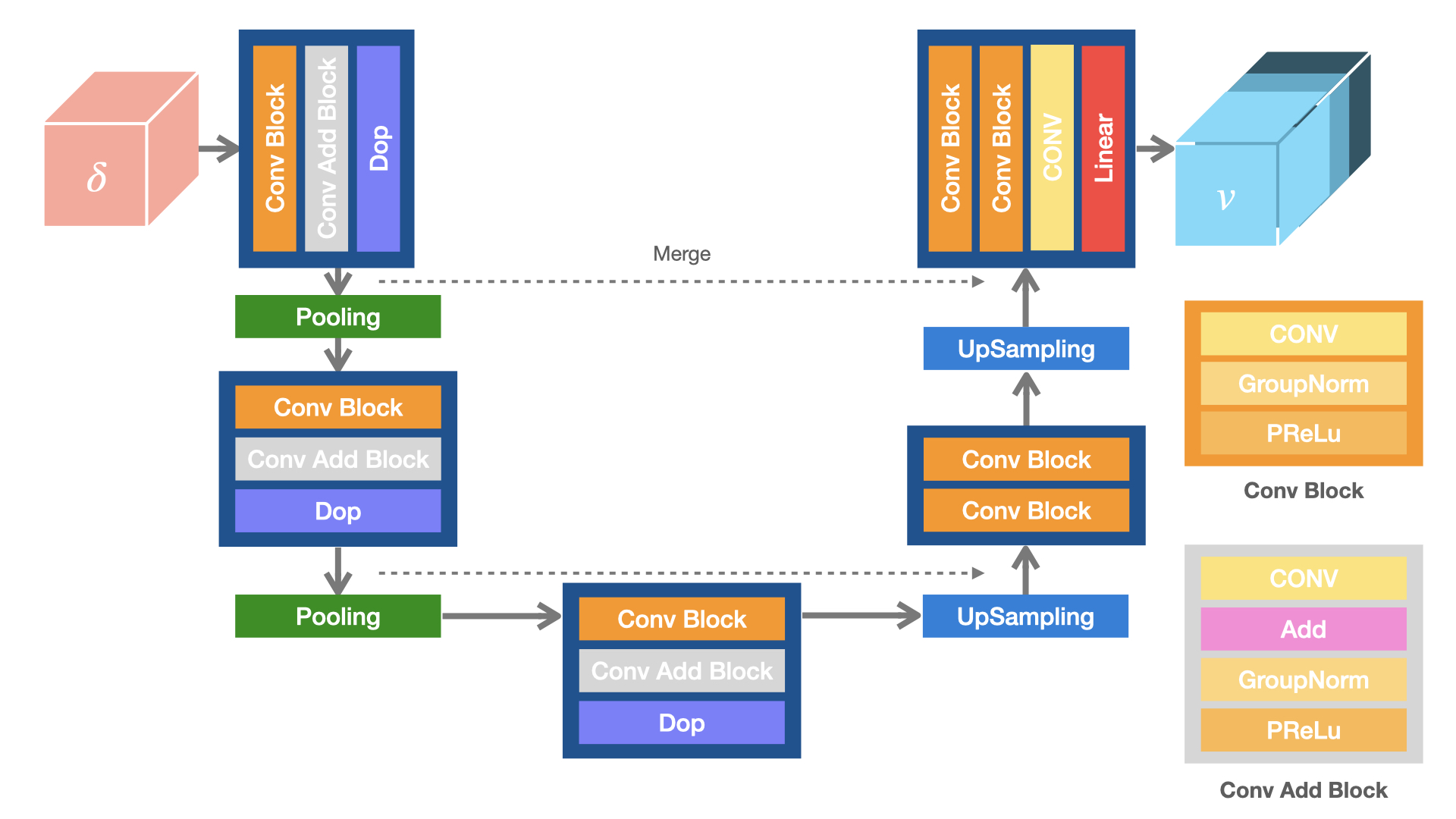}
\caption{\label{fig:Unet_3D} The Unet model architecture designed for reconstructing 3D peculiar velocity fields from 3D over-density fields.}
\end{figure*}

The deep learning algorithm, a subset of the machine learning methods characterized by its depth of layers, is built upon artificial neural networks that mimic the interconnections of neurons in the human brain. When reconstructing velocity fields from density fields, the data dimension expands from one channel to three. To achieve this dimensional expansion in the reconstruction, we adapt a 3D conventional U-net, which is recognized as a fully convolutional network. The architecture of our modified Unet model is illustrated in Fig.~\ref{fig:Unet_3D}.

The input over-density field is first processed through several downward blocks, including convolutional, GroupNormalization, activation, add, pooling and dropout layers, to compress features. The compressed feature then goes through several upward blocks, which are expanded by upsampling layers and previous features, to generate the output velocity field in three channels. Here are short descriptions of each layer:

\begin{itemize}
    \item \textbf{Convolutional layers} consist of numbers of kernels that are used to extract morphological features from the input matrix. High-level features are extracted at the initial Convolutional layer, while more abstract features are obtained later.

    \item \textbf{GroupNormalization layers} normalize the input by dividing the input channels into groups, which is more accurate and stable than the batch normalization.                       

    \item \textbf{Activation layers} provide activation functions for the corresponding Convolutional layers, which defines a node's output by acting on a linear combination of the input. We implement PReLu and Linear Activation functions in the architecture.

    \item \textbf{Add layers} add several input matrices together to one output matrix.
    
    \item \textbf{Pooling layers} reduce the spatial size of the input matrix. We implement Maxpooling in this paper, which reduces the input size by taking the maximum value over the input windows.
    
    \item \textbf{Dropout layers} systematically re-initialize a subset of neurons during each training epoch, which reduces the risk of over-training. The dropout rate is 0.2 in our model. Notably, the dropout layer is exclusively incorporated into the Unet-SDSS model, distinguishing it from the Unet model which does not include this dropout layer.
    
    \item \textbf{Upsampling layers} resize the input and expand its spatial size.
\end{itemize}

Training a neural network for 3D vector field reconstructions poses a challenge in selecting loss functions that significantly influences the accuracy of predictions. We implement a customized loss function that combines mean square error (MSE) loss, dice loss, and a velocity-weighted MSE loss, which can be expressed by $avg(\sum\limits_i v_t^i\sum\limits_i(v_t^i-v_p^i)^2)$, where $v_t^i$ is the component of the true peculiar velocity along $i$th direction and $v_p^i$ indicates the corresponding component of the predicted velocity.

\section{Simulation and Training Data} \label{sec:simul}

The deep learning algorithm constructs a data-driven model to predict anticipated features, demanding a substantial dataset for effective training. To Ensuring optimal performance of a deep learning model, the training data necessitates a high degree of reliability. Therefore, a majority of deep learning models in cosmology utilize simulated training data. In this paper, the training data is derived from the Halo catalog of the Outer Rim Simulation \citep{HeiFinPop2019}, which is a dark matter only simulation with cosmological parameters similar to those of the WMAP-7 \citep{KomSmiDun2011}, as outlined in Table~\ref{T_Mill}.

\begin{table}[!ht]
\caption{The cosmological parameters of the Outer Rim Simulation}
\centering
{
\begin{tabular}{lc}
\hline
Matter density, $\Omega_m$ & 0.2648\\
Cosmological constant density, $\Omega_\Lambda$ & 0.7352 \\
Baryon density, $\Omega_b$ & 0.0448\\
Hubble parameter, $h$ (100 km s$^{-1}$ Mpc$^{-1}$) & 0.71\\
Amplitude of matter density fluctuations, $\sigma_8$ & 0.8\\
Primordial scalar spectral index, $n_s$ & 0.963\\
Box size ($h^{-1}$Mpc) & 3,000\\
Number of particles & $10,240^3$\\
Particle mass, $m_p$ ($10^{9} h^{-1} M_{\odot}$) & 1.85\\
Softening, $f_c$ ($h^{-1}$kpc) & 3\\
\hline
\end{tabular}%
}
\label{T_Mill}
\end{table}

We segment the Outer Rim Simulation box (3000 $h^{-1}$Mpc) into 1,000 non-overlapping smaller boxes ensuring no correlation between them, each with dimensions of 300 $h^{-1}$Mpc. For each of these small boxes, we calculate the mass-weighted over-density field by halos with a grid size of 5 $h^{-1}$Mpc, while the velocity field is determined by the distribution of dark matter particles. Our model takes the 3D over-density field as input data and the 3D velocity field as output, both calculated using the nearest grid point method. In addition, the Cloud-in-Cell method and the Subtracted Dipole scheme are also commonly employed in calculating density fields. In this paper, the gridding scale effectively mitigates the distinctions arising from these various methods. We treat the halo as a Dirac delta function rather than a top-hat sphere, which has negligible difference from each other with large smoothing scales as discussed in \cite{WanMoYan2012}. Furthermore, varying types of input data, either the over-density field or the density field, negligibly impacts the performance of the deep learning model. To align with the requirements of the analytical method, we use the over-density field as the input data of the deep learning model.

\begin{figure*}
\centering
\includegraphics[width=17cm]{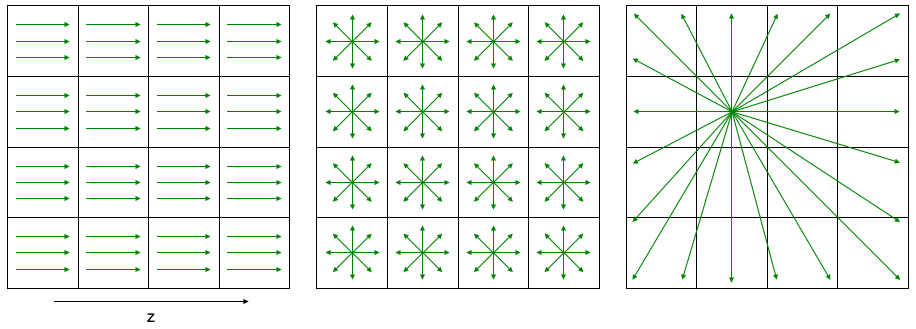}
\caption{\label{fig:RSD_setting} Illustration of three patterns for incorporating RSD into the training data. The green arrows indicate the direction of the RSD effect. The left panel displays the pattern where RSD is added along z-axis, while the middle panel shows the radial RSD pattern, and the right panel demonstrates a random RSD pattern where RSD is added along the radial direction of a specific small box.}
\end{figure*}

In consideration of the practical application of the deep learning model to observational data, we include Redshift Space Distortion (RSD) into the training data, which means calculating the deep learning model's over-density field in redshift space instead of real space. Further details about the data utilized for the deep learning model and the analytical approach is available in Table~\ref{T_Data}. The RSD is added to the training data in three patterns, as shown in Fig.~\ref{fig:RSD_setting},  to encompass a broader range of data samples. Firstly, the RSD along z-axis (left panel) is disigned to accommodate distant data samples, whose redshift distance is defined by $z_s = z + v_z/100$ in unit $h^{-1}$ Mpc, and $x_s = x$ and $y_s = y$, where the subscript $s$ indicates the redshift space. Secondly, the center-radial RSD (middle panel) is more suitable for nearby data samples, which is defined by $p^i_s = p^i + p^i*\frac{ (\textbf{r} - \textbf{r}_c) \cdot \textbf{v} }{ 100 |\textbf{r} - \textbf{r}_c| }$, where $p^i$ indicates the coordinate position of each halo along the $i$th direction (x, y and z), $\textbf{v}$ represents the vector of the peculiar velocity, $\textbf{r}$ is the distanct vector of halos and  $\textbf{r}_c$ indicates the distant vector of the center of each small box. Finally, a random RSD (right panel),  introduced by applying RSD along the radial direction of a specific box, is employed to access universality. It is expressed by $p^i_s = p^i + p^i*\frac{ (\textbf{r} - \textbf{r}_t) \cdot \textbf{v} }{ 100 |\textbf{r} - \textbf{r}_t|}$, where $\textbf{r}_t$ indicates the center coordinate of the specific box. In this paper, we primarily focus on the last pattern, since the model performances using different RSD patterns exhibit similarly. Notably, all the RSD patterns are applied to simulation halos, any values calculated from dark matter particles are unaffected by RSD effect to establish an ideal environment for the analytical method.

\begin{table}[!ht]
\caption{Input and output data of each model and method. The input source denotes the origin from which the over-density field is calculated, while the output source indicates the data source used for calculating the velocity field. The "IF RSD" columns illustrate whether the input or output data includes RSD effect.}
\centering
{
\begin{tabular}{lcccc}
\hline
     & \multicolumn{2}{c}{Input} & \multicolumn{2}{c}{Output}\\
\hline
     & Source & If RSD & Source & If RSD \\
Unet & DM Halos & $\surd$ & DM particles &  $\times$ \\
FFT  & DM particles & $\times$ & DM particles & $\times$ \\

\hline
\end{tabular}
}
\label{T_Data}
\end{table}

\section{Training}\label{sec:training}

The 1000 simulation boxes are randomly divided into training and validation sets, with an 80\%-20\% split, responsively. The model is initially trained using input data derived from dark matter halos at snapshot $z=0$, without imposing any mass threshold. To assess adaptability, the model is subsequently tested under different conditions: 1) same conditions as the training data, where the testing data mirrors the validation data; 2) using dark matter halos with a mass threshold $M_h>10^{12} h^{-1} M_{\odot}$ at snapshot $z=0$; 3) using dark matter halos in snapshot $z=0.21$ without mass threshold. Further details are listed in Table~\ref{T_Test}.

\begin{table}[!ht]
\caption{Training and testing data for the Unet model. The term "DM halos" refers to dark matter halos utilized in generating the input over-density field. The symbol "V" indicates that the validation data is also employed as testing data.}
\centering
{
\begin{tabular}{l@{\hskip 0.2in}ll}
\hline
        & Data                             &  Samples\\
\hline
\hline
Training& DM halos                               &  800\\
\hline
Validation& DM halos                             &  200\\
\hline
Test& DM halos                                   &  200 (V)\\
    & DM halos ($M_h>10^{12} h^{-1} M_{\odot}$)  &  1000\\
    & DM halos (z = 0.21)                        &  1000\\

\hline
\end{tabular}%
}
\label{T_Test}
\end{table}

The Unet model is trained on a NVIDIA Tesla V100 GPU, which takes about 50 minutes for 100 epochs with 16 batch size using 800 boxes training data and 200 boxes validation data. The primary challenge in the model training arises from the limit of GPU memory. Due to the data size and associated training parameters, the Unet model demands large memory during training. Consequently, adjusting the box size may be necessary when reducing the grid size.

\begin{figure}
\centering
\includegraphics[width=8.5cm]{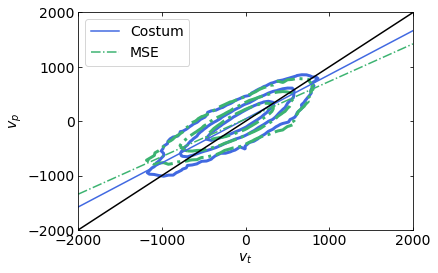}
\caption{\label{fig:Loss_compare} The magnitude of all velocity components in the reconstructed velocity field using different loss functions. The x- and y-axes represent the magnitudes of the true ($v_t$) and predicted ($v_p$) velocity components ($v_x$, $v_y$, $v_z$). The blue solid contours show the results of the deep learning algorithm with the custom loss function, and the green dash-dotted contours indicate the results using mean square error (MSE) loss function. The black solid line represents the 1:1 ideal relation between the true and predicted velocities. The solid blue and dash-dotted green lines are the linear fittings of the Unet and FFT results, respectively.}
\end{figure}

In our training experiments, we observed significant performance improvements by replacing the BatchNormalization layer with the GroupNormalization layer. Additionally, the Conv Add Block contributes to enhanced the model performance. For our Unet model, the custom loss function discussed in Section \ref{sec:method} outperforms both the mean square error (MSE) and variational autoencoder (VAE) loss functions. Fig~\ref{fig:Loss_compare} illustrates the distinct performance between the MSE and our custom loss function. Obviously, the model precision improves after incorporating the dice loss and the velocity-weighted MSE. However, a more comprehensive loss function could potentially lead to overfitting. To mitigate this, we recommend including dropout layers in the Unet model and adjusting the learning (or decay) rate based on the specific model and input data. In our model, the training loss exhibits a slight decreasing trend after the validation loss has converged. However, the difference between these losses is negligible with our epoch setting (100 epochs), as confirmed by the results using independent testing data in the subsequent sections.

Furthermore, the Unet model exhibits flexibility in accommodating various types of input training data. we conduct tests using following input data variations: 1) over-density field calculated by the mass-weighted halo distribution (referred to as mass over-density field); 2) over-density field calculated by the halo distribution (referred to as number over-density field); 3) density field calculated by the mass-weighted halo distribution (referred to as mass density field); 4) density field calculated by the halo distribution (referred to as number density field). The Unet model performs consistently across these different input data types, indicating that the deep learning algorithm can extract the linear density-velocity relation from density distribution features independently of their feature form or average magnitude. In the following discussion, we primarily employ the mass over-density field as the input data to align with the analytical method (section~\ref{sec:results}). However, for the SDSS data, we implement the mass and number density fields for groups and galaxies, respectively, due to the low number density in the SDSS survey (section~\ref{sec:observ}).

\section{Results} \label{sec:results}

We evaluate the performance of our deep learning model by comparing its prediction results with that of the analytical method. We also investigate the stability of our deep learning model under different conditions.

\subsection{Comparison with analytical method} \label{sec:DL_FFT}

\begin{figure*}
\centering
\includegraphics[width=8cm]{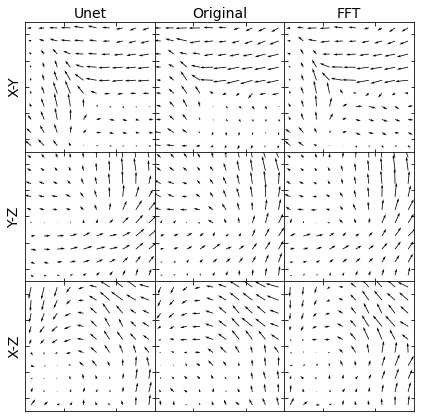}
\includegraphics[width=8cm]{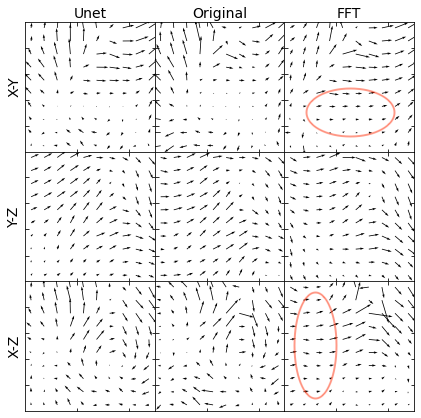}
\caption{\label{fig:No_Mass_limit} The velocity fields reconstructed through deelp learning algorithem (Unet) and analytical method (FFT) without mass threshold, alongside the true velocity field from simulations (Original). The left panel shows the results of a 50 $h^{-1}$Mpc central cutoff range of the box, while the right panel displays the boundary cutoff results of the box.}
\end{figure*}

In this subsection, we compare the performance of the deep learning model (Unet) with the analytical model (FFT) and the true velocity field (Original) calculated from the simulation, as depicted in Fig~\ref{fig:No_Mass_limit}. The Unet and FFT results are reconstructed from the over-density field, but the difference is that the over-density input of the Unet is calculated in redshift space by halos in the z=0 snapshot without any selections, while the over-density input of the FFT is generated in real space by dark matter particles in the same snapshot, as listed in Table~\ref{T_Data}. The Original result in the figure is the expected velocity field that is calculated directly by the velocities of dark matter particles.

The left panel of Fig~\ref{fig:No_Mass_limit} shows X-Y, Y-Z, and X-Z slices of a central (50 $h^{-1}$Mpc) cutoff out of the box (300 $h^{-1}$Mpc). In this figure, the Unet shows no significantly different performance than the analytical method in the central region of the box. However, in the boundary region (50 $h^{-1}$Mpc) of the box (right panel), the Unet outperforms the analytical method, especially for the nonlinear scales. Although performed under ideal conditions, the analytical method smooths out the nonlinear flows and provides inaccurate vector directions in the boundary region of the box, as shown by the red-circled parts in the right panel. In contrast, Unet captures more accurate nonlinear features.

\begin{table*}[!ht]
\caption{Prediction statistics of each model and method. The value in this table is averaged across all the testing boxes.}
\centering
{
\begin{tabular}{l@{\hskip 0.2in}c@{\hskip 0.2in}c@{\hskip 0.2in}c@{\hskip 0.2in}c}
\hline
Models           &  Fitting slope  & Residuals ($<200$ km/s) & correlation coefficient ($cos\theta$) & Average coherence ($\Gamma$) \\
\hline
FFT &                   0.70       &  77.6\%   &  0.70   &     0.62\\
Unet&                   0.81       &  87.7\%   &  0.83   &     0.79\\
Unet-SDSS-galaxy&       0.85       &  71.4\%   &  0.69   &     0.78\\
Unet-SDSS-group &       0.68       &  71.3\%   &  0.64   &     0.62\\

\hline
\end{tabular}
}
\label{T_Prediction}
\end{table*}

\begin{figure}
\centering
\includegraphics[width=8.5cm]{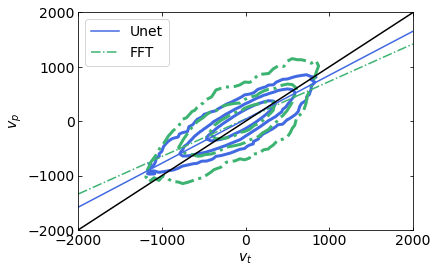}
\caption{\label{fig:velocity_scatter_noM} The magnitude of all the velocity components in the reconstructed velocity field for a random testing box. The x- and y-axes represent the magnitude of the true ($v_t$) and predicted ($v_p$) velocity components ($v_x$, $v_y$, $v_z$). The blue solid contours show the results of the deep learning algorithm (Unet), and the green dash-dotted contours indicate the results of the analytical method (FFT). The black solid line represents the 1:1 ideal relation between the true and predicted velocities. The solid blue and dash-dotted green lines are the linear fittings of the Unet and FFT results, respectively.}
\end{figure}

\begin{figure}
\centering
\includegraphics[width=8.5cm]{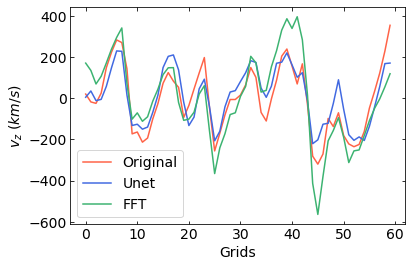}
\caption{\label{fig:v1d_value_noM} $v_z$ component values of a randomly selected row along $z$ direction. The y-axis shows the $v_z$ value, and the x-axis indicates the grids along $z$ direction of the testing box, which is not displayed in the figure. The blue line indicates the result of the deep learning algorithm (Unet), the green line shows the result of the analytical method (FFT), and the red line illustrates the original true value (Original).}
\end{figure}

\begin{figure}
\centering
\includegraphics[width=8.5cm]{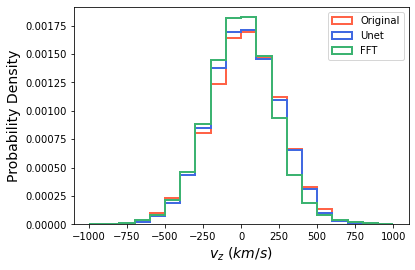}
\caption{\label{fig:v1d_value_PDF} The probability density of the $v_z$ component in a random selected testing box. The y-axis shows the $v_z$ value, and the x-axis indicates the normalized probability density. The blue line indicates the result of the deep learning algorithm (Unet), the green line represents the result of the analytical method (FFT), and the red line illustrates the original true value (Original).}
\end{figure}

Besides direction, magnitude is another crucial factor of a vector field. Fig.~\ref{fig:velocity_scatter_noM} compares the predicted magnitude ($v_p$ on x-axis) of velocity components of the velocity field reconstructed by Unet and FFT with the true component magnitude ($v_t$ on y-axis) of the Original velocity field. Both methods underestimate the velocity magnitude, but Unet's result has a fitting slope of 0.81, which is better than FFT's slope of 0.70. Additionally, Unet's result shows tighter scatter contours than FFT's result. For a more obvious contrast, Fig.~\ref{fig:v1d_value_noM} shows the $v_z$ component values of a randomly selected row in a testing box. In the figure, both the Unet and FFT results show general agreement with the Original result. However, FFT's result has a very obvious overestimation near the boundary region. This kind of unexpected large overestimation is not rare in FFT's prediction and may be caused by the sub-optimal performance of the analytical method in the boundary region. Fig.~\ref{fig:v1d_value_PDF} displays the probability density of the $v_z$ component values of the testing box. The Unet predictions demonstrate a close alignment with the true expectations across most $v_z$ bins, while the FFT's $v_z$ probability density deviates from the expectations, which reinforces the superiority of the deep learning algorithm over the analytical method. 

\begin{figure*}
\centering
\includegraphics[width=8cm]{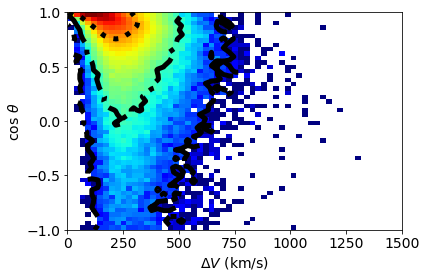}
\includegraphics[width=8cm]{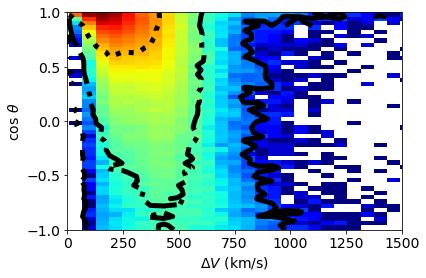}
\caption{\label{fig:Vector_compare} Magnitude and direction differences between the predicted velocities and the true velocities. The left panel shows the difference between the Unet predictions and the Original true velocities, while the right panel represents the FFT predictions. The x-axis indicates the magnitude difference and the y-axis displays the direction difference. The dotted, dash-dotted and solid contours represent the 68\%, 95\%, and 99.7\% data regions, respectively.}
\end{figure*}

Fig.~\ref{fig:Vector_compare} combines the magnitude and direction information to evaluate the performance of the Unet (left panel) and FFT (right panel) with the original velocity field. The x-axis displays the magnitude difference ($\Delta V = | \textbf{v}_p - \textbf{v}_t |$), and the y-axis represents the direction difference ($cos\theta = \frac{\textbf{v}_p \cdot \textbf{v}_t}{|\textbf{v}_p| |\textbf{v}_t|}$). The Unet's result exhibits tighter contours than the FFT's result at all levels of 68\%, 95\%, and 99.7\%, indicating that the Unet performs better than the FFT in vector field reconstruction. Notably, The $cos\theta$ can be regarded as a geometric correlation coefficient for velocity vectors, expressing the correlation between the true and predicted velocities. The correlation coefficient is about 0.83 for the Unet model, while this value is only about 0.70 for the FFT result, which underscores that the Unet model outperforms than the analytical method. Remarkably, our Unet model produces vector matrices as output, posing challenges in calculating standard deviations which is required in the conventional correlation coefficient. Therefore, we implement the geometric correlation coefficient as an alternative to the conventional correlation coefficient in this paper.

Additionally, a comparison between the predictions of the deep learning algorithm and the analytical method is conducted using the average coherence,  which is given by $\Gamma = avg(\frac{\textbf{v}_p\cdot\textbf{v}_t}{\textbf{v}_t\cdot\textbf{v}_t})$, where $\textbf{v}_p$ indicates the predicted peculiar velocity, $\textbf{v}_t$ is the true velocity, and $avg()$ denotes the average function. The Unet model demonstrates an average coherence of $\Gamma_{Unet}=0.79$, outperforming the FFT result with an average coherence of $\Gamma_{FFT}=0.62$. The average prediction statistics of all the testing boxes are listed in Table~\ref{T_Prediction}.

\begin{figure}
\centering
\includegraphics[width=8.5cm]{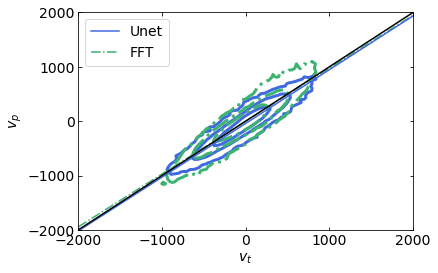}
\caption{\label{fig:velocity_scatter_noM_center} Similar to Fig.~\ref{fig:velocity_scatter_noM}, but showing results of the 200 $h^{-1}$ Mpc central cutoff of a random testing box.}
\end{figure}

As discussed earlier, the sub-optimal performance of the analytical method might be caused by the edge effect. To investigate this, we selected an example box and compared the performance of the Unet and FFT methods with a central 200 $h^{-1}$ Mpc cutoff, as shown in Fig.~\ref{fig:velocity_scatter_noM_center}. For the central part of the box, the FFT result has a slope equal to 0.97, which is very close to the slope of the Unet result (0.99). However, not all testing boxes have results as good as this example. For those boxes that have cells with very large velocity magnitudes, both the Unet model and FFT analytical method provide underestimated predictions.

Consequently, the Unet and the FFT yield comparable results in the central region of the box, whereas the deep learning algorithm significantly outperforms the analytical method in the boundary region with simpler calculation processes. Additionally, the deep learning model is adaptable to different conditions, which will be discussed in the next subsection.

\subsection{adaptability} \label{sec:DL_adapt}

Due to observation conditions, only bright and large enough galaxies can be accurately detected. Therefore, using all the halos in the simulation box would not fit the observation conditions. To emulate the observational situation, we apply a minimum mass threshold of $10^{12}$ $h^{-1} M_{\odot}$ to the simulation halos and use the selected halos to generate over-density fields for the velocity reconstruction. For the analytical method, the halo mass threshold would introduce bias to the velocity reconstruction, which requires an additional parameter, referred as halo bias, to mitigate the effect. For the deep learning algorithm, this problem can be solved by training the model with input data selected by the same mass threshold. However, for the purpose of our study, we employ the model trained on data without mass selection when applied to the mass-selected testing data. This choice allows us to evaluate the Unet's adaptability to datasets of varying mass selection conditions, as outlined in Table~\ref{T_Test}.

\begin{figure}
\centering
\includegraphics[width=8.5cm]{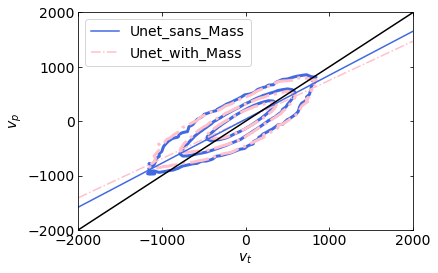}
\caption{\label{fig:velocity_scatter_M12} Similar to Fig.~\ref{fig:velocity_scatter_noM}, but displaying Unet results only.
The blue solid contours show the results of the testing data without mass threshold, while the pink dash-dotted contours indicate the results of the testing data with a mass threshold of $10^{12}$ $h^{-1} M_{\odot}$.}
\end{figure}

Fig.~\ref{fig:velocity_scatter_M12} shows the results obtained by applying the same Unet model, initially trained by data of all halos without mass threshold, to two types of testing data: 1) data without mass threshold (blue); 2) data subject to a mass threshold that selects halos larger than $10^{12}$ $h^{-1} M_{\odot}$ (pink). To accentuate this effect, we use the same example box as in Fig.\ref{fig:velocity_scatter_noM}, ensuring consistency in comparison. Therefore, the blue contours in this figure are identical to those in Fig.\ref{fig:velocity_scatter_noM}, since the model and testing data are the same. The pink contours show underestimated predictions when the model is applied to testing data with a minimum mass threshold, resulting in a slope of 0.72. Despite differences in training and testing conditions, the Unet consistently outperforms the FFT, indicating that the deep learning algorithm is less sensitive to the halo mass conditions than the analytical method. However, the deep learning model's performance diminishes when confronts the mismatch between the training and testing conditions. Training the model with data that agrees with the testing condition would maximize the effectiveness of the deep learning algorithm. As a result, training the model with simulation data that mimics the observational condition would be an optimal choice.

\begin{figure}
\centering
\includegraphics[width=8.5cm]{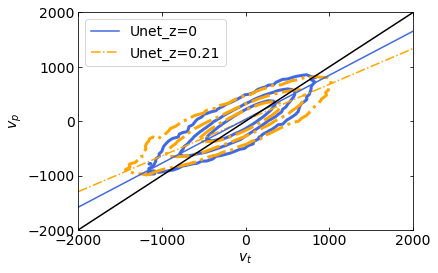}
\caption{\label{fig:velocity_scatter_Z021} Similar to Fig.~\ref{fig:velocity_scatter_noM}, but representing Unet results only.
The blue solid contours show the results of the testing data in the $z=0$ snapshot, and the orange dash-dotted contours indicate the results of the testing data in the $z=0.21$ snapshot.}
\end{figure}

In simulations, the time evolution is represented by discrete snapshots, which differs from the continuous redshift changing in real observations. Therefore, a model trained by data from a specific snapshot may not perform consistently well in a deep observational survey. To test the adaptability of the deep learning algorithm to various redshift data, we apply the Unet model trained with data from $z=0$ snapshot to testing data from different snapshots, as exemplified in Fig.~\ref{fig:velocity_scatter_Z021}. In the figure, the blue contours display the results of applying the $z=0$ Unet model to $z=0$ testing data, while the orange contours indicate predictions of applying the same model to $z=0.21$ testing data, resulting in a slope of 0.66. Obviously, applying the Unet model to a dataset from a different snapshot may lead to less accurate predictions. Furthermore, the degree of the underperformance improves as the redshift difference between the training and testing snapshots decreases, but worsens when the redshift difference increases. In such scenarios, it is necessary to train the Unet model separately with data corresponding to different redshifts in order to suit the model to distinct redshift regions within a deep survey.

\section{Observation Applications} \label{sec:observ}

\begin{figure}
\centering
\includegraphics[width=8.5cm]{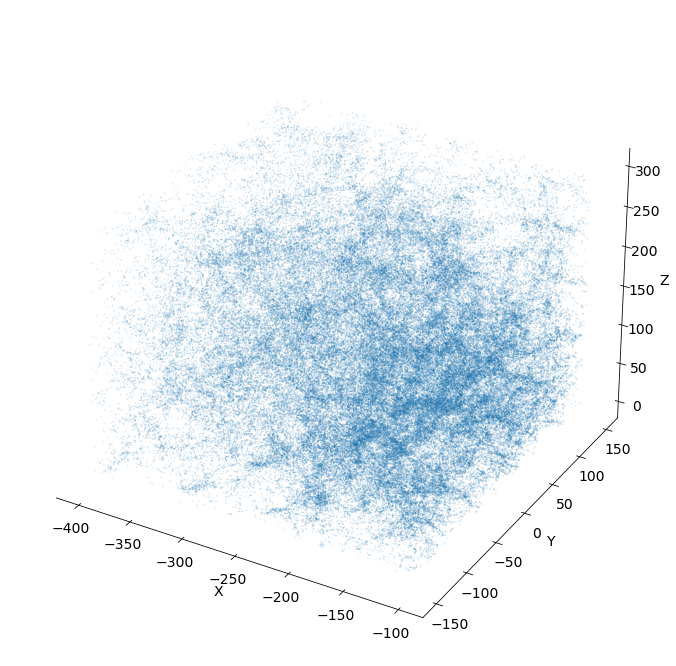}
\caption{\label{fig:SDSS_group} The group distribution in a 300 $h^{-1}$ Mpc box from the SDSS-group catalog.}
\end{figure}

To evaluate the performance of the Unet model on observations, we conduct tests with the Sloan Digital Sky Survey \cite[SDSS,][]{SDSS2000}. We extracted a 300 $h^{-1}$ Mpc cubic region from the SDSS Data Release 7 \cite[SDSS DR7,][]{SDSSDR7}, which comprises approximately 210,900 galaxies from the SDSS-galaxy catalog and 141,700 groups from the SDSS-group catalog \citep{YanMoBos2007, YanMoBos2012}, as illustrated by the group distribution in Fig.~\ref{fig:SDSS_group}. Notably, the distribution of SDSS groups (or galaxies) is not uniform; rather, it has a radial distribution pattern where the number of groups (or galaxies) decreases along the x and z axes. 

As discussed in Subsection~\ref{sec:DL_adapt}, optimizing the effectiveness of the Unet model requires training data closely aligns with the observational conditions. To generate mock catalogs appropriately, Firstly, we characterize selection function of the SDSS box by calculating the number of galaxies (or groups) in redshift bins along radial direction (radial redshift distribution). Secondly, each of the 1000 small simulation boxes is shifted to match the coordinates of the SDSS box, as indicated by the x, y and z axis in Fig.~\ref{fig:SDSS_group}. Finally, we generate 1000 SDSS galaxy (or group) catalogs by randomly selecting halos larger than $10^{11}$ $h^{-1} M_{\odot}$ to match the SDSS galaxy (or group) selection function. It is essential to highlight that the number density of SDSS-galaxy (or SDSS-group) catalog is significantly lower than the halo number density in the simulation. The number density of SDSS-galaxy (or -group) mock catalog is about 7.8 (or 5.2) $\times 10^{-3}$ $[h^{-1}$ Mpc$]^{-3}$, while the halo number density of the simulation box is about 68.6 $\times 10^{-3}$ $[h^{-1}$ Mpc$]^{-3}$. Compared to the training data implemented in Section~\ref{sec:results}, the training data for the Unet-SDSS model has a significantly lower number density, which reflects the poor observational conditions. Given the low number density of the SDSS mock catalogs, we utilize density as the input data of the Unet-SDSS model instead of the over-density implemented in Section~\ref{sec:results}.

The training data for the Unet-SDSS model is derived from the SDSS mock catalogs. As the SDSS galaxy catalog lacks mass information, whereas the SDSS group catalog provides mass of each group, we bifurcate the Unet-SDSS model into two models: Unet-SDSS-galaxy and Unet-SDSS-group. For the Unet-SDSS-galaxy model, the input is the number density field calculated from the SDSS galaxy mock catalogs, with the output (velocity filed) derived from the dark matter particles, as detailed in Table~\ref{T_Data_SDSS}. For the Unet-SDSS-group model, we implement the mass density field extracted from the SDSS group mock catalogs as the input and velocity field from dark matter particles as the output.

\begin{table*}[!ht]
\caption{Input and output data for the Unet-SDSS model. The ``Type" column specifies the type of the input data, with $\rho_N$ representing the number density field and $\rho_M$ indicating the mass density field. The input source identifies the implemented source for calculating the density field, while the output source denotes the data source used for the velocity field. The ``IF RSD" columns illustrate whether the input or output data includes RSD effect.}
{
\centering
\begin{tabular}{l@{\hskip 0.5in}ccc@{\hskip 0.5in}cc}
\hline
     & \multicolumn{3}{c}{Input} & \multicolumn{2}{c}{Output}\\
\hline
                 &    Type  & Source            & If RSD  & Source      & If RSD \\
Unet-SDSS-galaxy & $\rho_N$ & SDSS galaxy mocks & $\surd$  & DM particles & $\times$ \\
Unet-SDSS-group  & $\rho_M$ & SDSS group mocks  & $\surd$ & DM particles & $\times$ \\

\hline
\end{tabular}
}
\label{T_Data_SDSS}
\end{table*}

Similar to the Unet model, each of the Unet-SDSS models is trained using 800 out of the 1000 SDSS mock boxes, and is tested by the remaining 20\% of the boxes. The training strategy aligns with the approach discussed in Section~\ref{sec:training}.

\begin{figure}
\centering
\includegraphics[width=8.5cm]{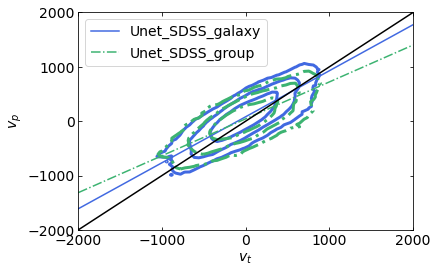}
\caption{\label{fig:SDSS_scatter} The magnitude of all the velocity components in the Unet-SDSS model's reconstructed velocity field for a random testing box. The x- and y-axes show the magnitude of the true ($v_t$) and predicted ($v_p$) velocity components ($v_x$, $v_y$, $v_z$). The blue solid contours indicate the results of the Unet-SDSS-galaxy model, and the green dash-doted contours display the predictions of the Unet-SDSS-group model. The black solid line represents the 1:1 ideal relation between the true and predicted velocities. The blue solid line and the green dash-dotted line are the linear fittings of the Unet-SDSS-galaxy and Unet-SDSS-group models, respectively.}
\end{figure}

Fig.\ref{fig:SDSS_scatter} shows simulation examples of Unet models trained under SDSS like conditions. Compared with the Unet model trained without observation-specific conditions, the Unet-SDSS models exhibit diminished performance evidenced by larger scatters. For the Unet-SDSS models, only about 71\% of the result has residual magnitude smaller than 200 km/s, while this value increases to about 87\% for the Unet model discussed in Section\ref{sec:results}. The underperformance of the Unet-SDSS models is likely caused by the low number density and the unbalanced distribution, which leads to biased density fields. This can be improved with larger datasets in further observation surveys. Under the SDSS observational conditions, the model performs better while using the SDSS-galaxy mock, which has a slope equal to 0.85 (blue line), than using the SDSS-group mock with a slope equal to 0.68 (green dash-dotted line). The different performance between the Unet-SDSS-galaxy and -group models indicates that a larger data sample provides more accurate predictions, even though the group model uses mass density field as input data which carries more information than the galaxy model's input data, the number density field. \cite{GanLilNus2023} explores the impact of galaxy number density on the velocity reconstruction neural network, revealing that the MSE loss reduces with an increase in the number density \citep[See Figure 13 in][]{GanLilNus2023}. It further explains the superior performance of the SDSS-galaxy model than the SDSS-group model. Details regarding the predictions of each model are listed in Table~\ref{T_Prediction}.

Table~\ref{T_Prediction} presents the average statistics for the Unet-SDSS models. Across all the statistical parameters (fitting slope, residuals, correlation coefficient and average coherence), the Unet-SDSS-galaxy model consistently outperforms the Unet-SDSS-group model, highlighting the significant influence of the sample number density on the Unet performance. Notably, the average coherence of the Unet-SDSS-galaxy model is significantly surpasses that of the analytical method, however, its correlation coefficient is comparable to the analytical method. While both the correlation coefficient and the average coherence quantify the model performance, they yield distinct outcomes. In contrast to the conventional correlation coefficient, the correlation coefficient implemented in this paper exclusively considers the geometric correlation between predictions and expectations. On the other hand, the average coherence includes both the magnitude and the geometric information of vectors. Consequently, the average coherence offers a more precise explanation of model performance in this study.

For observational data like SDSS, the redshift distribution of galaxies or groups is a biased probe of the true mass distribution, which induces challenges into the analytical velocity reconstruction. Therefore, a density reconstruction step preceding the velocity reconstruction in analytical methods becomes necessary, which makes the velocity reconstruction of observational data computationally complex and intensive. In contrast, deep learning algorithms provide a simple and direct approach to reconstruct velocity field from redshift surveys without complicated data preprocessing. 

\begin{figure*}
\centering
\includegraphics[width=16cm]{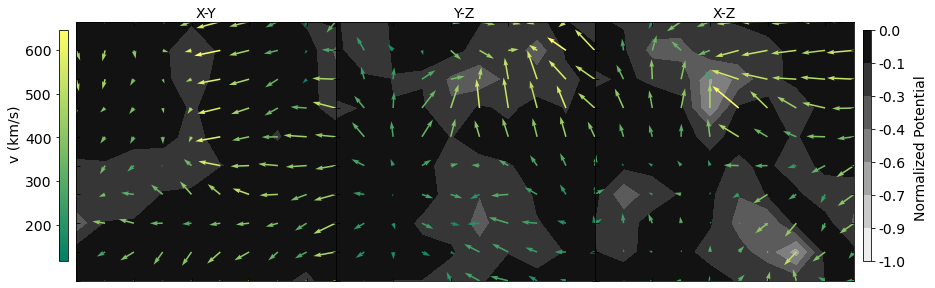}
\caption{\label{fig:SDSS_field} The 3D velocity fields reconstructed by the Unet-SDSS-galaxy model of a 100 $h^{-1}$ Mpc central cutoff. The colorful arrows shows the reconstructed velocity field of the Unet-SDSS-galaxy model, along with the normalized gravitational potential field calculated from dark matter particles, representing the true potential field. Note: The figure has been mean-pooled with a kernal size of two for better visualization.}
\end{figure*}

Fig.~\ref{fig:SDSS_field} explains the Unet model's capacity of reconstructing peculiar velocity field from incomplete and biased density field. The figure shows the reconstructed velocity field of the Unet-SDSS-galaxy model through colorful arrows. The grey contours represent the normalized gravitational potential field derived from the dark matter particles within the simulation box, representing the true and unbiased gravitational potential. Although the true potential field is not well reproduced by the number density field of the SDSS galaxy mocks, the reconstructed velocity field are consistent with each other and closely aligns with the true potential field. That is, even if biased input data are given, the deep learning algorithm automatically corrects the bias and provides accurate outputs that match with the expectations. The spatial plot of the Unet-SDSS-group model is omitted from this paper as it exhibits no significant difference from Fig.~\ref{fig:SDSS_field}. The potential contour is visualized more distinctly in the figure compared to the density contour, where the gradient is less discernible than that of the gravitational potential. Therefore, we represent the spatial contour using the gravitational potential in the figure.

\begin{figure*}
\centering
\includegraphics[width=18cm]{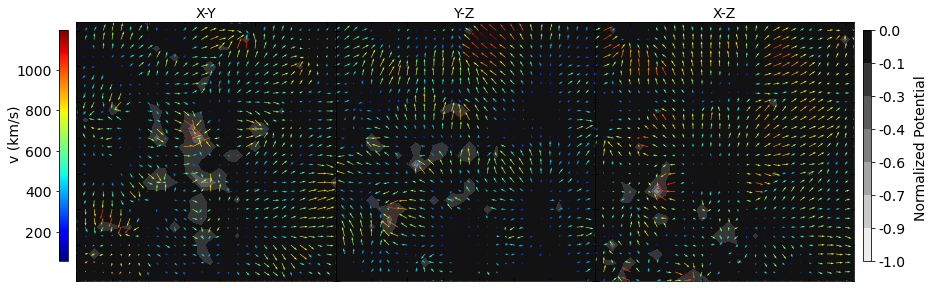}
\includegraphics[width=18cm]{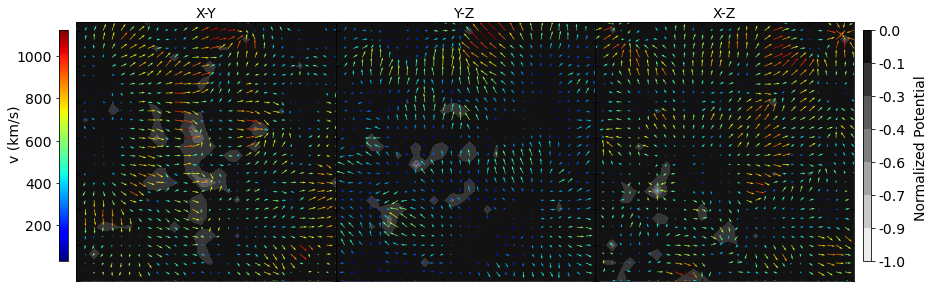}
\caption{\label{fig:SDSS_Observ} The 3D reconstructed velocity field of the 300 $h^{-1}$ Mpc SDSS-galaxy and SDSS-group catalog box. The upper panel shows the reconstructed velocity field of the SDSS-galaxy catalog using the Unet-SDSS-galaxy model, along with the normalized potential field derived from the mass density field of the SDSS-group catalog. The lower panel displays the reconstructed velocity filed of the SDSS-group catalog using the Unet-SDSS-group model with the same potential field as in the upper panel. Note: The figure has been mean-pooled with a kernal size of two for better visualization.}
\end{figure*}

Finally, the Unet-SDSS model is applied to the SDSS observation data, and the reconstructed velocity field of the 300 $h^{-1}$ Mpc SDSS box is shown in Fig.~\ref{fig:SDSS_Observ}. The upper panel displays the reconstructed velocity field of the SDSS-galaxy catalog using the Unet-SDSS-galaxy model, and the lower panel shows the reconstructed velocity field of the SDSS-group catalog using the Unet-SDSS-group model. In both panels, the grey contours represent the identical normalized gravitational potential field derived from the mass density field of the SDSS-group catalog. In this figure, the reconstructed velocity field partially matches with the SDSS potential field. This is because the group mass distribution is not an unbiased tracer to generate true potential field. Furthermore, the velocity fields reconstructed through SDSS-galaxy and -group catalogs show differences from each other. By comparing the prediction statistics of the galaxy and group models using simulation data in Table~\ref{T_Prediction}, we expect that the galaxy catalog would be a more optimal input for the SDSS velocity reconstruction due to its larger sample size. Considering the agreement between the reconstructed velocity field and the true potential field in the simulation test, we anticipate that the reconstructed velocity field of this SDSS observation data is suitbale for further statistical studies.

\section{Conclusion}
\label{sec:conclusion}

The analytical method used to reconstruct the peculiar velocity field from obsevational redshift surveys involves a series of intricate procedures, including density reconstruction, halo bias correction, and iterative calculations. In this paper, we build a Unet model with deep learning algorithms to directly reconstruct 3D pecuilar velocity field from the redshift survey data with a simplified and enhanced process.

The Unet model is initially trained and tested using simulations, and its performance is compared to that of an analytical method conducted under an ideal condition. The analytical method utilizes the over-density field of dark matter particles as input data, which is isolated from any bias that might be caused by halo mass distribution. In contrast, the Unet is implemented under more realistic input conditions, using halo mass over-density fields with RSD effect. Despite the running condition being much more restricted than that of the analytical method, the Unet model provides predictions with enhanced precision and reduced scatters, particularly for the boundary region of the data box. Furthermore, the Unet model shows capability of extracting non-linear features in velocity reconstructions, which are usually smoothed out in the analytical result.

Halos are biased tracers of the true mass distribution, which introduces bias into velocity reconstructions when using halo mass over-density as input data. Additionally, the bias varies with different minimum halo mass thresholds. To address this, the analytical method employs a halo bias parameter for correction. In this paper, we evaluate the Unet model's adaptability with varying minimum halo mass thresholds and observe that it exhibits a lower sensitivity to the halo mass compared to the analytical method. However, the various halo mass thresholds still impact the model performance. To solve this issue, we recommend training the Unet model using the same halo mass threshold as the testing data.

One challenge unique to the Unet model, but not the analytical method, is that the simulation cannot provide training data with continuous time evolution, which leads to discrete redshift between different snapshots. Therefore, a Unet model trained on data within a specific snapshot may not be suitable for all the regions of a deep redshift survey, while the analytical method does not need to deal with this issue. We test the performance of the Unet model trained in the $z=0$ snapshot on different testing data from various snapshots, and find that the larger the redshift difference between the training and testing shapshots, the more significant the bias in the testing results. As a remedy, it is advisable to train the Unet model separately with different snapshots to accommodate the diverse redshifts in different regions of a deep survey.

The results of our tests on the halo mass threshold and the snapshot redshift emphasize the importance of training the Unet model with simulation data that emulate the observational conditions for optimal performance. To achieve this, we generate mock catalogues (SDSS mocks) that mimic the SDSS DR7 data to train the Unet model for SDSS observational velocity reconstructions. The Unet-SDSS model is first tested by the SDSS mocks. Due to the lower number density and non-uniform distribution of galaxies (or groups) in the SDSS dataset, the Unet-SDSS model's result has larger scatters and residuals from the true expectations. However, it is noteworthy that despite these challenges, 71\% of the Unet model's residual remain below 200 km/s. In addition, the reconstructed velocity field from the Unet-SDSS model closely agrees with the true potential field of the dark matter particles. Considering the large uncertainty in peculiar velocity measurements, we expect that the velocities reconstructed by the Unet-SDSS model are suitable for peculiar velocity statistics. In the end, we apply the Unet-SDSS model to the SDSS observation data and display the reconstructed SDSS peculiar velocity field.

In conclusion, using deep learning algorithms to reconstruct 3D peculiar velocity fields from redshift surveys is feasible and promising. It simplifies the analytical velocity reconstruction method and yields more precise results.

\begin{acknowledgments}
This work is supported by the National Science Foundation of China (Nos. 11833005, 11890692), 111 project No. B20019, and Shanghai Natural Science Foundation, grant No.19ZR1466800. We acknowledge science research grants from the China Manned Space Project with No.CMS-CSST-2021-A02.
\end{acknowledgments}

\bibliographystyle{aasjournal}
\bibliography{Yuyu}

\end{document}